# Is Journal Citation Indicator a good metric for Art & Humanity Journals currently?

Yu Liao[1], Li Li[2] and Zhesi Shen[3]

[1] liaoyu@mail.las.ac.cn
National Science Library, Chinese Academy of Science, Beijing, China

[2] lili2020@mail.las.ac.cn
National Science Library, Chinese Academy of Science, Beijing, China

[3] shenzhs@mail.las.ac.cn
National Science Library, Chinese Academy of Science, Beijing, China

**Abstract**

Probably Not.
Journal Citation Indicator (JCI) was introduced to address the limitations of traditional metrics like the Journal Impact Factor (JIF), particularly its inability to normalize citation impact across different disciplines. This study reveals that JCI faces significant challenges in field normalization for Art & Humanities journals, as evidenced by much lower correlations with a more granular, paper-level metric, CNCI-CT. A detailed analysis of Architecture journals highlights how journal-level misclassification and the interdisciplinary nature of content exacerbate these issues, leading to less reliable evaluations. We recommend improving journal classification systems or adopting paper-level normalization methods, potentially supported by advanced AI techniques, to enhance the accuracy and effectiveness of JCI for Art & Humanities disciplines.

**Introduction**

The Journal Impact Factor (JIF) has long been the predominant metric for evaluating journals, celebrated for its simplicity and widespread acceptance (Miles et al., 2018). However, its limitations have been widely criticized, including its failure to account for variations in citation potential across disciplines (Althouse et al., 2009; Nederhof 2006), differences in document types, the constraints of a short citation window, and the impact of highly skewed citation distributions (Larivière and Sugimoto, 2019; Bordonset al., 2002). To address some of these issues, the Journal Citation Indicator (JCI) was introduced. JCI calculates the average Category Normalized Citation Impact (CNCI) of articles published in a journal, normalized using a journal-level subject category classification system (hereafter referred to as JCI-WoS).

In recent years, JCI has gained traction as a metric, especially for evaluating Art & Humanities journals, which face unique challenges due to their distinctive citation practices and field-specific characteristics (Torres-Salinas et al., 2022). Beginning with the 2023 Journal Citation Reports (JCR), Clarivate Analytics adopted JCI-based quartiles, replacing JIF-based quartiles, for Art & Humanities journals. Although prior studies have demonstrated a strong correlation between JCI and JIF for journals indexed in SCIE and SSCI, the performance of JCI as a field-normalization metric for Art & Humanities journals remains insufficiently examined.

This study evaluates JCI's effectiveness for Art & Humanities journals by comparing it with CNCI calculated based on Citation Topics (hereafter referred to as CNCI-CT), a more granular, paper-level classification system. A high correlation between JCI and CNCI-CT indicates that JCI has effectively achieved field normalization. Specifically, this study addresses the following research questions:

1. Does the correlation between JCI and CNCI-CT for Art & Humanities journals differ from that observed for Science and Social Science journals?
2. If differences exist, what underlying factors contribute to these discrepancies?

Through this investigation, we aim to provide a detailed evaluation of JCI's field-normalization performance in Art & Humanities journals, offering valuable insights into its appropriateness as a standard metric for these disciplines.

**Data and Methods**

To evaluate the performance of the JCI in the context of Art & Humanities journals, we obtained the CNCI values of 22,979 journals indexed in SCIE, SSCI, AHCI, and ESCI from the InCites database in December 2024. Only documents categorized as articles and reviews published during the 2021–2023 period are included. For each journal, two CNCI values were extracted:

1. **CNCI based on Subject Category (JCI):** This is calculated at the journal level by normalizing the citation impact of articles against all other documents within the same journal's subject category as defined by the Web of Science. JCI is essentially the average CNCI-WoS for a journal.
2. **CNCI based on Citation Topics-meso level (CNCI-CT):** This is calculated at the paper level by normalizing citation impact based on a more granular, hierarchical classification system called Citation Topics. The meso level topics is selected as it has similar granurity with subject category.

**Results**

**RQ1:** Does the correlation between JCI and CNCI-CT for Art & Humanities journals differ from that observed for Science and Social Science journals?

**A1:** Low correlations between JCI and CNCI-meso are found for Art & Humanity related subject categories, implying JCI's in-effectiveness in citation field normalization.

Figure 1 illustrates the correlation coefficients between JCI and CNCI-CT across various subject categories. A higher correlation indicates a closer alignment between the two metrics for journals within a given category. We grouped subject categories into three broad groups: Science, Social Science, and Art & Humanities.

As shown in Figure 1, the Science and Social Science groups exhibit consistently high and tightly clustered correlation coefficients, reflecting strong alignment between JCI and CNCI-CT. In contrast, the Art & Humanities group displays a wider range of lower correlation coefficients. Notably, categories such as *Architecture* and Theater show the weakest correlations. This discrepancy indicates that there is a significant divergence between journal-level classification-based and paper-level classification-based normalization metrics for journals in the Arts & Humanities, raising concerns about the effectiveness of JCI in normalizing citation disparities within these subject categories.

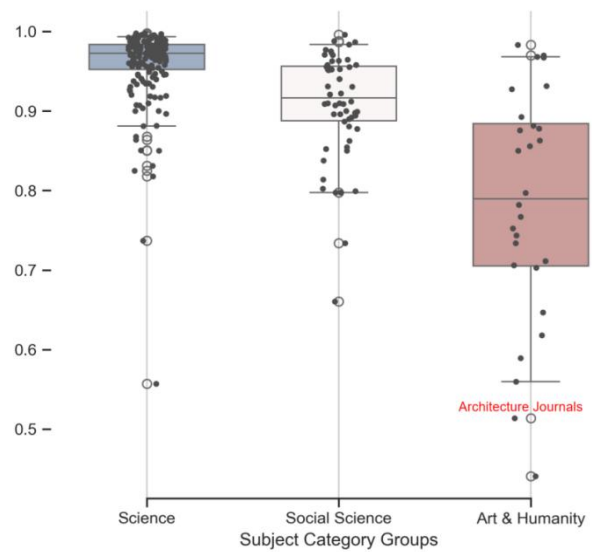

**Figure 1. Correlation coefficients between JCI and CNCI-CT for subject categories.**
Each dot represents a subject category, with its value indicating the correlation coefficient. Categories are grouped into three clusters: Science, Social Science, and Art & Humanities.

**RQ2:** If differences exist, what underlying factors contribute to these discrepancies?
**A2:** Through the case study of architecture journals, it was revealed that some art & humanities journals, despite publishing a significant number of science or social science papers, are not co-assigned to the science and social science categories. This omission results in these journals gaining a substantial advantage in the JCI.

Among Art & Humanities categories, *Architecture* demonstrates one of the lowest correlations between JCI and CNCI-CT. To further investigate, we analyzed the structure of this category in detail.

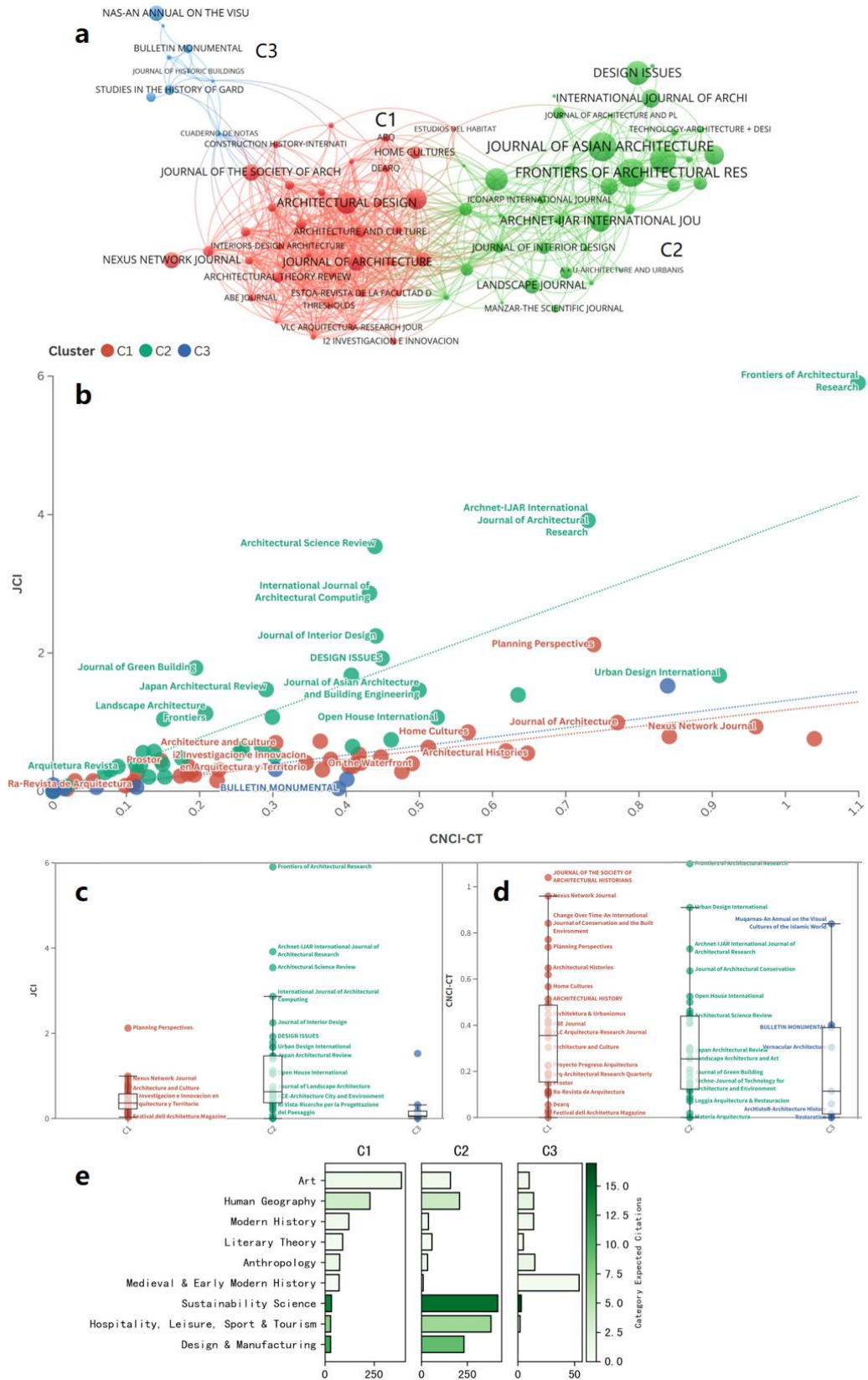

**Figure 2. Field normalization performance for Architecture journals.** (a) Similarity network of Architecture journals, showing three identified clusters. (b) Scatter plot comparing JCI and CNCI-CT values for journals in each cluster. (c) Distribution of JCI values by cluster.

(d) Distribution of CNCI-CT values by cluster. (e) Distribution of covered citation topics of each cluster with color representing the category expected citation.

*Cluster Analysis.* Figure 2(a) presents the similarity network of Architecture journals, where journals publishing similar content are positioned closer together. Using a community detection algorithm, we identified three distinct clusters.

*Disparities in JCI and CNCI-CT Across Clusters.* Figures 2(c) and 2(d) compare the distributions of JCI and CNCI-CT values across the three clusters. As shown in Figure 2(c), Cluster 2 (green dots) exhibits significantly higher JCI values compared to Clusters 1 and 3. However, Figure 2(d) reveals that CNCI-CT values are more evenly distributed across all three clusters. The scatter plot in Figure 2(b) highlights the substantial advantage that JCI provides to journals in Cluster 2, suggesting that JCI does not fully account for citation disparities within the *Architecture* category.

*Content Differences Across Clusters.* An examination of publication topics in Cluster 2 journals reveals a higher proportion of articles related to *sustainability science* topics with higher citation potential compared to traditional Architecture topics, as shown in Fig.2(e). Despite this interdisciplinary content, most Cluster 2 journals remain solely classified under the *Architecture* category, with only a small number being co-classified into science or social science categories.

*Implications for Field Normalization.* Because JCI uses journal-level subject category normalization, Cluster 2 journals benefit significantly from their inclusion in a single, less-cited category, despite publishing content that overlaps with higher-citation Science fields. In contrast, CNCI-CT employs paper-level normalization based on Citation Topics, which more effectively captures thematic and disciplinary diversity, resulting in a more balanced evaluation of journals across clusters.

**Conclusions and Discussions**

This study examines whether the Journal Citation Indicator (JCI) effectively addresses field normalization challenges for Art & Humanities journals. By comparing JCI with CNCI-CT, a field-normalized indicator based on paper-level classification, we find significantly lower correlations between the two metrics in Art & Humanities categories. This indicates that JCI currently struggles to handle field normalization disparities in these fields.

A detailed analysis of *Architecture* journals reveals that this issue primarily arises from journal-level misclassification. Similar patterns are observed in other Art & Humanities categories, such as *Art* and *Religion*. Due to the lower citation density characteristic of Art & Humanities compared to Science and Social Science fields, the effects of misclassification are more pronounced, further reducing the reliability of JCI in these areas.

To address these limitations, we recommend prioritizing the optimization of journal classifications (Yu et al., 2025) before expanding the use of JCI. Alternatively, adopting a paper-level classification system for field normalization (Sichao et al., 2023) could provide a more robust solution. However, implementing paper-level classification in Art & Humanities faces unique challenges: approximately 20% of papers in these fields are not assigned citation topics, compared to nearly 0% in Science and Social Science. To overcome these challenges, advanced AI methods, including large language models (LLMs), could be employed to assign citation topics based on titles and abstracts. These tools have the potential to improve classification coverage significantly, enhancing the accuracy of field normalization and making metrics like JCI more reliable for evaluating Art & Humanities journals.